\documentclass[runningheads]{llncs}
\usepackage[T1]{fontenc}
\usepackage{graphicx}
\usepackage[numbers]{natbib}  
\usepackage{url} 
\usepackage{lipsum}

\newcommand\blfootnote[1]{%
  \begingroup
  \renewcommand\thefootnote{}\footnote{#1}%
  \addtocounter{footnote}{-1}%
  \endgroup
}
\begin{document}
\title{Effect of external characteristics of a virtual human being during the use of a computer-assisted therapy tool}
%
%

\author{Navid Ashrafi\inst{1,2*}\orcidID{0009-0005-8398-415} \and
Vanessa Neuhaus\inst{1*} \and
Francesco Vona\inst{1}\orcidID{0000-0003-4558-4989} \and
Nicolina Laura Peperkorn\inst{3}\orcidID{0009-0008-9481-9354} \and
Youssef Shiban\inst{3}\orcidID{0000-0002-6281-0901} \and
Jan-Niklas Voigt-Antons\inst{1}\orcidID{0000-0002-2786-9262}}
\authorrunning{Ashrafi et al.}
%
\institute{University of Applied Sciences Hamm-Lippstadt, Marker Allee 76-78, 59063 Hamm, Germany\\
\email{navid.ashrafi@hshl.de}\\
\email{neuhausvanessa@web.de}\\
\email{francesco.vona@hshl.de}\\
\email{jan-niklas.voigt-antons@hshl.de}\and
Technical University of Berlin, Straße des 17. Juni 135, 10623 Berlin, Germany 
\email{ashrafi@tu-berlin.de}\\
\and Private University of Applied Sciences Göttingen, Weender Landstraße 3-7, 37073 Göttingen, Germany\\
\email{shiban@pfh.de}\\
\email{peperkorn@pfh.de}\\
$*$ Navid Ashrafi and Vanessa Neuhaus contributed equally to this publication.
}
\maketitle              
\blfootnote{This paper has been accepted for publication in the Human-Computer Interaction International Conference 2024. The final authenticated version is available online at https://doi.org/[will be added after the release of the paper].}
\thispagestyle{empty}
\pagestyle{empty}

\begin{abstract}
The relevance of identification within media and its capacity to impact the user effectively has been a research focal point for years whether it is a real or fictional character. Identification in the media context shapes behavior and broadens the user's social and emotional experience.
In immersive media (such as video games), virtual entities, e.g., virtual agents, avatars, or Non-Player Characters (NPCs), bridge the gap between users and the virtual realm. The sense of immersion is usually accompanied by a higher degree of identification. When using visual representations, i.e., in the form of an avatar or agent, new challenges arise concerning the visual design. In this context, media effects, especially identification, can again make the interaction more pleasant and attractive. While in many contexts of use, decisions can be made regarding visualization using the target group, research in recent years points to the potential of user-defined design, so-called customization. Although it seems obvious here that users visualize a virtual image of themselves in such cases, there are also other approaches to using customization. An essential question in this context is whether and to what extent the identification with a virtual avatar could influence the user experience of a psychological intervention. In this work, we explore the effect of visual similarity of a virtual anthropomorphic agent on the user experience in an intervention to reduce the effect of dysfunctional beliefs. In an experiment, 22 participants were asked to create a virtual agent in two groups, similar and dissimilar to them, and then, the avatar confronted them with their dysfunctional thoughts. The results show that the similarity of the virtual agent is not only associated with statistically significant increased identification but also with a positive influence on emotions and intrinsic motivation (more interest and enjoyment). This work contributes to the exploration of customization and identification, especially with virtual agents, and the potential implications for their visual design in the context of computer-assisted therapy tools.

\end{abstract}

\section{Introduction}

In media studies, extensive research spanning decades has focused on the crucial role of identification and its effective utilization to influence consumers \cite{Cohen2001, hoffner2005}. Whether manifested through real or fictional personalities, presented in text, image, or sound, identification within the media landscape is widely acknowledged for its transformative impact on behavior, expanding consumer social perspectives, and enriching emotional experiences \cite{Cohen2001, jansz2005, klimmt2009, konijn2007}.

While the influence of identification in traditional media has been extensively explored, an expanding area of interest lies in immersive media, such as video games or serious games. On these interactive platforms, consumers assume a more active role. Virtual characters serve as direct interfaces between users and the virtual world, with the immersive nature often correlating with heightened levels of identification \cite{hefner2007}. This immersive effect extends beyond entertainment media to interactive applications in fields like medicine, especially in developing computer-assisted tools to diagnose, treat, prevent, and rehabilitate mental illnesses.

The exploration of novel therapeutic approaches is particularly relevant given the increasing global prevalence of mental health issues \cite{who2022}. Computer-assisted tools present several potential advantages, such as time and cost savings, increased acceptance, accessibility, availability, and reduced barriers for patients seeking assistance. In this context, virtual avatars and agents act as natural interfaces, representing users or therapists and assuming roles such as assistants, caregivers, or interview partners \cite{bualan2020, meeker2016, mitruț2021}. Examples in cognitive behavioral therapy include the text-based chatbot Woebot studied by Fitzpatrick et al. \cite{fitzpatrick2017} and the application Help4Mood by Burton et al. \cite{burton2016}, featuring a 3D visualization of a virtual agent.

The use of visual representations, such as avatars or agents, in these applications introduces new challenges in terms of visual design. Media effects, notably identification, can be intentionally leveraged to enhance interaction, making it more natural, pleasant, and appealing. Recent research emphasizes the potential of user-defined design, commonly called customization \cite{birk2019, birk2018, ratan2016}. While it may seem intuitive for users to visualize a virtual representation of themselves, other approaches, such as individual creation, have been explored. Avatar therapy by Leff et al. \cite{leff2014} is an example where patients with schizophrenia give a face to imaginary voices through a virtual anthropomorphic avatar.

A similar concept was pursued in a recent pilot study targeting mental illnesses like depression and anxiety disorders \cite{kocur2021}. Rooted in Beck's cognitive therapy \cite{beck1979}, patients engage in a dialogue with an avatar to challenge dysfunctional beliefs with alternative, functional thoughts \cite{kocur2021}. Although the personal customization of avatars was initially excluded in this study, discussions of the results hinted at the potential therapeutic benefits of allowing users to modify the avatar's appearance. However, another study by Pimentel and Kalyanaraman \cite{pimentel2020}, exploring the visualization of negative self-concepts, suggests that increased identification following personal customization of the avatar could be negatively influenced when exposed to negative stimuli.

Hence, a pivotal question arises concerning how the degree of identification, particularly in visualizing negative self-concepts with a virtual avatar or agent, may impact the user experience of computer-assisted therapy tools addressing dysfunctional cognitions \cite{kocur2021}. This question serves as the focal point of this paper.

To address the research question on the impact of visual similarity on user experience in computer-assisted therapy, five hypotheses have been derived to leverage existing scientific contributions. An experiment incorporating user customization features to design virtual agents will then be conducted. The potential effects, particularly those associated with increased identification, were examined using data collection instruments measuring emotional well-being and motivation. 

\textbf{RQ:} How does a virtual anthropomorphic agent's visual similarity or dissimilarity influence identification and user experience when interacting with a computer-assisted therapeutic tool addressing dysfunctional beliefs?

\textbf{H1:} The confrontation with a visually self-similar virtual agent has a higher negative impact on identification than a confrontation with a dissimilar agent.

\textbf{H2:} After confronting a visually self-similar virtual agent, the discrepancy between positive and negative emotional well-being is significantly higher than after interacting with a dissimilar agent.

\textbf{H3:} Confronting a visually self-similar virtual agent has a more significant negative impact on interest and enjoyment after customization than encountering a dissimilar agent.

\textbf{H4:} Following the confrontation with a visually self-similar virtual agent, the perceived value and usefulness are considered overall higher than interaction with a dissimilar agent.

\textbf{H5:} Confronting a visually self-similar virtual agent negatively impacts perceived pressure and tension after customization more than a dissimilar agent.


\section{Related Work}

\subsection{On Identification}

In psychology, identification encompasses both a defense mechanism and a process where individuals adopt traits of significant others, contributing to personality development \cite{laughlin1970}. It is primarily an unconscious process, although it can be partially preconscious or conscious (Laughlin, 1979; Schafer, 1973). Drawing on Freud \cite{freud1942}, Wollheim \cite{wollheim1974}, and Bettelheim \cite{bettelheim1943, bettelheim2010} theories, the identification process involves temporarily relinquishing one's identity awareness, allowing a person to view the world from another's standpoint \cite{Cohen2001}. Erikson \cite{erikson1968} notes its development in childhood, intensifying during adolescence with influences from peers and new authorities. It is part of healthy psychological development, fostering independence \cite{cramer1991}. In social cognitive learning theory, Bandura \cite{bandura1986, bandura2001} emphasizes identification's role in shaping behavior, influenced by perceived similarity and motivated by rewards. Notably, identification is not limited to humans and can extend to non-human entities \cite{schafer1968}. The consequences of identification, determined by consciously or unconsciously chosen models, can be positive, fostering self-esteem, self-transcendence, and a sense of meaning and belonging, or negative and destructive \cite{mael2001}.

Two additional types of identification, often applied in the media context, are similarity identification and wishful identification, shedding light on consumers' relationships with media characters \cite{Cohen2001, hoffner2005, konijn2007}. Similarity identification, rooted in Bandura's insight \cite{bandura1986}, emphasizes the significance of the perceived similarity between an individual and a model in predicting the replication of traits. It is synonymous with 'perceived similarity,' acknowledging the subjectivity of perceived resemblance rather than an objectively measurable one \cite{bandura1986}. In contrast, wishful identification involves the psychological process where an individual desires or attempts to resemble a model in appearance or behavior \cite{hoffner1996, hoffner2005}. The distinction lies in whether identification is based on existing, similar traits or those one wishes to incorporate into one's identity. Both forms, however, are closely interconnected, as similarity identification often triggers the desire to emulate another person or character, especially those perceived as popular or successful, such as media stars \cite{bandura2001, basil1996}.

\subsection{Virtual Agents and Avatars}

After providing a concise overview of identification and its associated theories, the focus shifts to defining another pivotal term. Distinguished from a virtual avatar, a virtual agent, also known as a conversational agent, simulates human conversations through text or oral language \cite{shaked2017}. In a broader context, it is a user interface facilitating system interactions with end users \cite{sciuto2018}. This interaction may be based on a predefined script, like a decision tree, or guided by artificial intelligence \cite{shaked2017}. Notable examples include voice assistants like Apple's Siri, Microsoft's Cortana, and chatbots used in customer service on online retailer websites. Virtual agents fall into two categories based on appearance—those without visual representation and embodied conversational agents \cite{cassell2001}, often portrayed in two or three dimensions. Depending on the chosen level of detail, embodied conversational agents can incorporate nonverbal elements like facial expressions and gestures, enhancing the interaction's naturalness \cite{louwerse2005}.

In contrast, a virtual avatar, like an agent, acts as an interface between a user and a digital application. Particularly in immersive media like video games, the critical distinction is that a virtual avatar always graphically represents a user in a virtual space, enabling interaction through controls like a gamepad \cite{gazzard2009}. They find applications in video games, social media, and other virtual spaces, representing individuals graphically \cite{gazzard2009}. Unlike virtual agents, virtual avatars always represent one or more users, controlled by the user. In immersive applications, such as video games or serious games, most scientific contributions assume real-time control of virtual avatars, in contrast to the interaction with a virtual agent. This distinction underscores the nuanced dynamics between virtual agents and avatars, influencing user experiences within digital environments.

\subsection{Identification in Immersive Media and Healthcare}

Virtual avatars, particularly in immersive contexts like video games, present a unique case in exploring identification. Unlike traditional media, the active role of consumers or players fosters a monadic relationship with a media character, which Kimmt et al. \cite{klimmt2009} also called “true" identification. This process involves a convergence of player and player character, leading to the adoption of feelings, goals, and perspectives \cite{klimmt2009}, which is also theorized to lead to the manifestation of a distinct self-concept during exposure \cite{bessiere2007, ducheneaut2009, klimmt2009, trepte2010}. This idea is rooted in Higgins' Self-Discrepancy Theory \cite{higgins1987} and implies that such convergence may lead to reduced self-discrepancy, especially when aspects of the ideal self of the player are present in the avatar \cite{bessiere2007, klimmt2009}. 
Recognizing the effects of identification in immersive media also extends from Cohen's work \cite{Cohen2001, cohen2006}. Positive outcomes include enhanced media enjoyment \cite{hefner2007, li2016, trepte2010} and increased persuasiveness of messages \cite{moyer2011, moyer2010}. Moreover, heightened identification positively correlates with intrinsic motivation, impacting user engagement across various applications like serious games or self-improvement tools \cite{birk2018, kao2018, oksanen2013}. 

Another aspect that has been given much attention in recent years is customization, which is also said to impact identification positively \cite{birk2018, birk2019, trepte2010}. Turkay \& Kinzer \cite{turkay2015} suggest that this is related to Self-Determination Theory \cite{ryan2000}, of which the main components are autonomy, connectedness, presence or immersion, and intuitive control \cite{rigby2006}. User similarity \cite{soutter2016}, personality traits \cite{soutter2016, dunn2012}, time invested \cite{turkay2015}, and narrative elements \cite{moyer2011} also foster identification, as well as the overall user motivation to get invested \cite{klimmt2009}. Understanding these factors is essential for designing applications that utilize identification for enhanced user experience and engagement.

While extensive research exists on identification in video games, the applicability of identification effects to digital applications in other domains, particularly medicine, remains a substantial area for further exploration. Computer-assisted therapy offers advantages regarding therapist time savings and self-administration by individuals \cite{wright2005, andersson2005}. Despite demonstrated effectiveness in treating symptoms like depression \cite{clarke2005, fitzpatrick2017, robinson2016}, adherence issues persist, significantly beyond controlled trials \cite{christensen2009, kelders2012}. Addressing this, intrinsic motivation from increased identification, as proposed by Birk et al. \cite{birk2018}, could prove beneficial. However, given variations in interactive elements, such as virtual avatars and agents, it is crucial to explore user relationships with virtual agents in different scenarios, as evidenced by existing research on virtual agent perception. Prior research has explored user perceptions of virtual agents, explored by studies on Woebot \cite{fitzpatrick2017}, where users anthropomorphized the virtual agent, referring to it as a "friend" or a "funny little guy". Various studies have also delved into how the appearance of agents influences user preferences \cite{baylor2009, baylor2004, khan2009}, acceptance in different contexts \cite{bickmore2010, micoulaud2016}, their role-model potential \cite{rosenberg2008, baylor2004}, and the impact of non-verbal behavior \cite{kulms2011, kruzic2020}.
In the study's chosen context of addressing dysfunctional cognitions, users confront their negative beliefs through a virtual agent \cite{kocur2021}. Like Pimentel and Kalyanaraman's \cite{pimentel2020} approach, users visualize and distance themselves from a negative self-concept. Exploring emotional responses to varied portrayals of one's negative self, this study investigates the effectiveness of confronting oneself to mitigate negative thoughts. Incorporating customization, based on prior research, enhances identification and potentially improves the application's personalization, motivation, and overall efficacy in treating dysfunctional cognitions. The study probes into how scenarios like similarity and dissimilarity may yield positive or negative effects, considering both intrinsic motivation and general well-being.
\section{Methodology}

\subsection{Data Collection}

\textbf{Identification with the Virtual Agent (PIS):} 
The identification between users and the virtual agent was measured using van Looy et al.'s \cite{vanLooy2010} scale for identification in online games. Specifically, the avatar identification subscale was employed, comprising six items each for similarity identification and embodied presence, and five for wishful identification. Group and game identification scales were excluded. The items were translated into German and slightly adjusted for contextual relevance, mainly by substituting "avatar" with "agent" while maintaining the original meaning. All items were rated on a 5-point Likert scale from 1 = "Strongly disagree" to 5 = "Strongly agree".

\textbf{Emotional Well-being (PANAS):} Given the potential for strong emotional reactions during the engagement with negative thoughts, they were also recorded using the German version of the Positive and Negative Affect Schedule (PANAS) \cite{breyer2016}, based on the English version by Watson, Clark, and Tellegen \cite{watson1988} which assesses emotional well-being with 20 items, ten describing positive affects and the other ten describing negative affects in adjectives. All items were rated on a 5-point Likert scale from 1 = "Very slightly or not at all" to 5 = "Extremely" This questionnaire was chosen because it shows moderate correlations with the negative effect of the Hopkins Symptom Checklist (HSCL), Beck Depression Inventory (BDI), and State Anxiety Scale (STAI) tests, commonly used for diagnosis in the medical context \cite{watson1988}.

\textbf{Intrinsic Motivation (IMI):} To measure intrinsic motivation, particularly linked to enjoyment in identification, the Intrinsic Motivation Inventory by Ryan and Deci \cite{ryan2000} was translated into German and applied. Using the interest/enjoyment, pressure/tension, and value/usefulness subscales, the assessment delved into the relationship between these values during customization and confrontation. Participants rated items on a 7-point Likert scale (1 = "Strongly disagree" to 7 = "Strongly agree"). This framework \cite{ryan2000} has been previously employed in a similar context \cite{birk2018}.

\textbf{Additional Data:} Alongside adapted frameworks, a supplementary questionnaire was created. Users provided general ratings for the therapeutic tool, the tested settings, the virtual agent, and the interaction. They evaluated the utility of customizing the virtual agent in a therapeutic context, expressed satisfaction with available options, and offered suggestions. Ratings were on a 7-point Likert scale, and settings were on a scale from 0 to 10.

\subsection{Experiment Setup}

\textbf{Therapy Tool:} The therapy tool used in the experiment was built based on the tool used in the original study \cite{kocur2021}. Confrontation with personal dysfunctional cognitions was facilitated through a virtual avatar controlled in real-time there, while the newly designed tool used a script-based virtual agent. To allow customization for the two different settings, a 3D model from DAZ 3D Studio \cite{daz2022}, was used. Customization options for the upper body of the virtual agent included variations in gender, body build, facial features, facial hair, hairstyle, and eye and skin color. General customization options were selected based on users' preferences that emerged from previous scientific contributions \cite{ducheneaut2009, turkay2015}. When designing customization options for the therapy tool, emphasis was placed on the virtual agent's external features rather than its personality traits, given the assumption that the agent naturally assumes a personality, particularly in confrontational situations. After preparation, the 3D model was embedded into the Unity game engine \cite{unity2021} and using the SALSA LipSync Suite \cite{salsa2022}, audio clips of negative thoughts, created online with text-to-speech in two variations, were played in real-time with matching animations. Non-verbal automatic behavior was previously tailored to enable the virtual agent to alternate between skeptical and angry facial expressions.
\begin{figure}[htbp]
\centerline{\includegraphics[width=10cm, height=6.2cm]{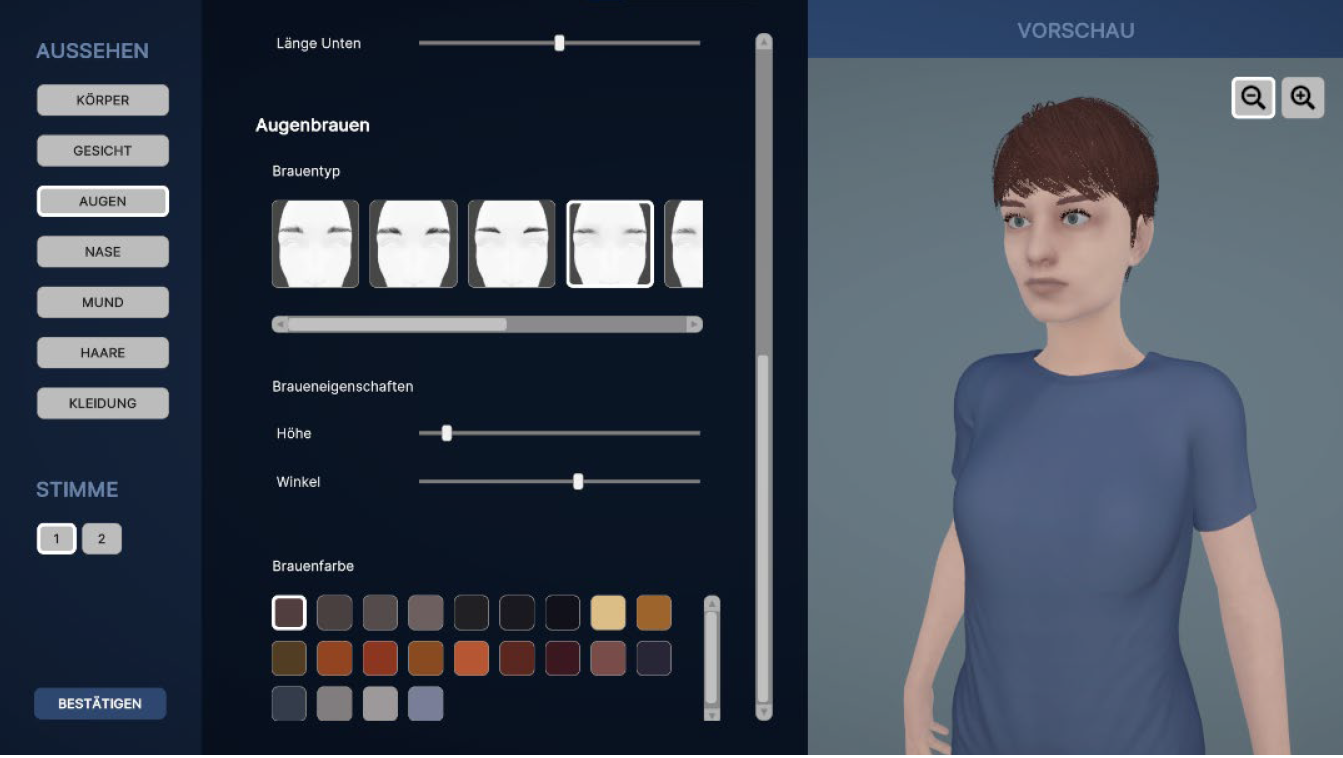}}
\centering
\caption{Customization scene setup.}
\label{comptence}
\end{figure}

\textbf{Experiment Procedure:} The experiment, approved by the Ethics Commission of the University of Applied Sciences Hamm-Lippstadt, was conducted at the Lippstadt campus in Germany over four days. Participants were briefed verbally and in writing about the experiment's content, purpose, and procedures before providing official consent. The experiment utilized a desktop computer and tablet for data collection across seven subsequent sessions. Participants recorded personal dysfunctional beliefs, and demographic data, and filled out questionnaires on dysfunctional attitudes using the Dysfunctional Attitude Scale (DAS-18 A) \cite{rojas2022} and self-esteem with the German version of Rosenberg's Self-Esteem Scale (RSES) \cite{vonGernot2003}.

After completing these questionnaires, participants engaged with the therapy tool. They customized the virtual agent and filled out questionnaires on identification, positive/negative affect, and intrinsic motivation. Following this, participants faced confrontation with the agent and, once again, filled out the same questionnaires presented after customization. They repeated this process in both, the similarity and dissimilarity settings. In the similarity setting, participants aimed to create an agent resembling them, while in the dissimilarity setting participants aimed to customize an avatar that would not necessarily look similar to them. The order of settings alternated for each participant. The entire experiment lasted approximately 45 minutes.

\subsection{Participants}

Participants \textit{(N = 22)} comprised nine females, twelve males, and one non-binary person. Their age ranged from 22 to 40 years, with an average age of 27 and they were comprised of diverse educational levels. On a scale of 1 = "Not at all" to 7 = "Very good," 20 participants rated their overall technology experience with a score of four or higher. Seven of them rated their previous experience as "Very good" with an average score of about 5.7. However, only four out of the 22 participants considered their experience as good, with an average score of 2.05. Concerning mental health, five users reported a history of psychotic symptoms, three were currently on therapeutic treatment, and three had received a diagnosis of a mental disorder. Additionally, six participants exhibited conspicuous values in self-esteem, all of which also showed high values in dysfunctional attitudes. Overall, 16 individuals demonstrated elevated values for dysfunctional attitudes, with an average score of 59.2, and for self-esteem, the average score was 19.

\section{Results}

\subsection{Identification}

To confirm that the similarity and the dissimilarity settings significantly differ regarding the degree of identification, they were compared using a t-test, both after customization (PRE) and after confrontation (POST). In this section, we will provide an overarching description of our t-test analysis and the implications for our five main hypotheses.


\subsection{Hypotheses}

\begin{figure}[htbp]
\centerline{\includegraphics[width=10cm, height=6.2cm]{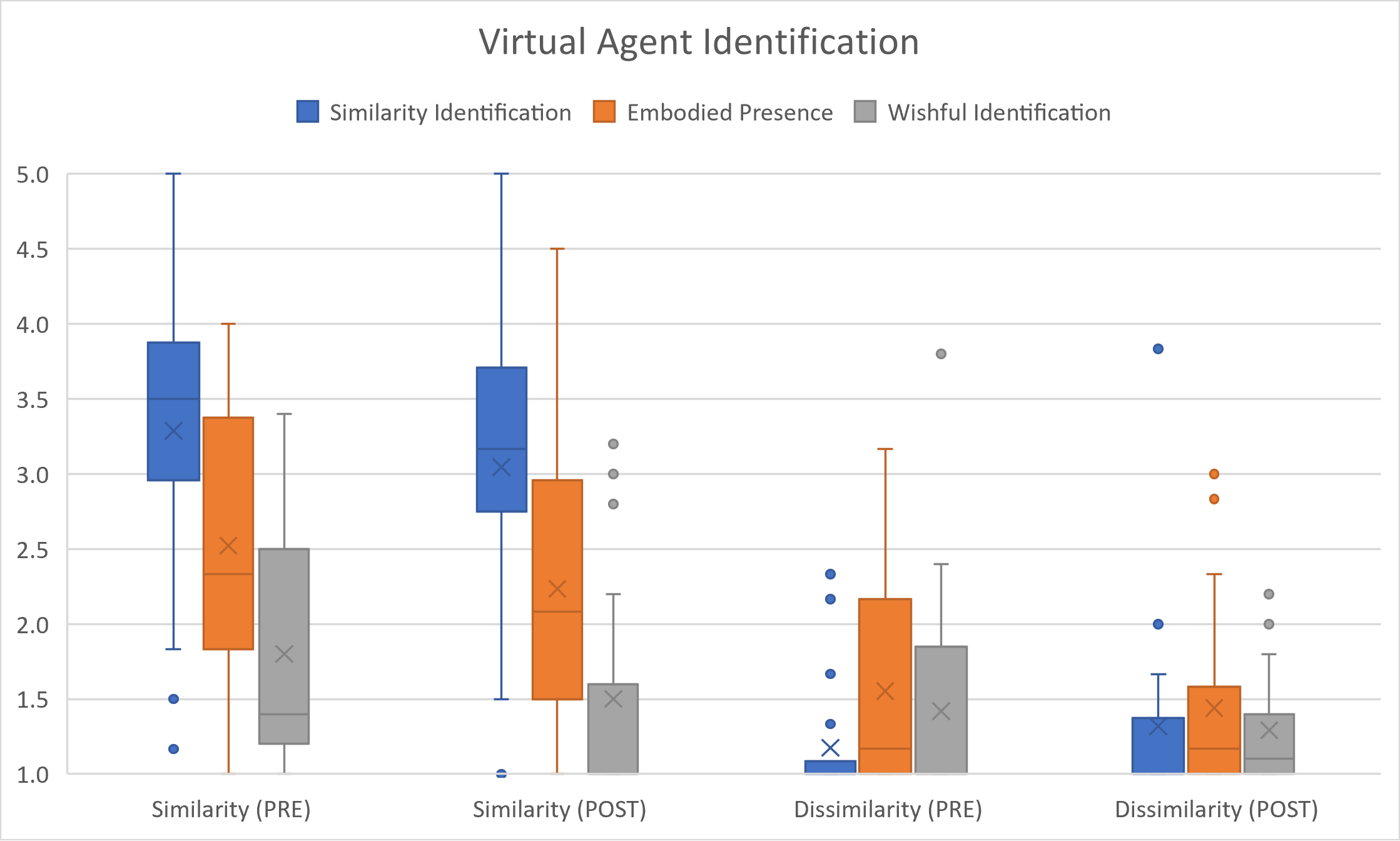}}
\centering
\caption{Identification with Virtual Agent after Customization (PRE) and after Confrontation (POST).}
\label{pre}
\end{figure}

\textbf{H1:} The initial hypothesis, pertaining to the extent of impact of the confrontation on overall identification within the contexts of similarity and dissimilarity, was substantiated through a t-test for dependent samples. In the similarity setting, the values for comparison before (M = 2.54, SD = 0.8) and after (M = 2.26, SD = 0.82) yielded t(21) = 2.23, p = .019, and d = 0.47, signifying statistical significance with a moderate effect size. Conversely, in the dissimilarity setting with customization (PRE) (M = 1.38, SD = 0.47) and confrontation (POST) (M = 1.35, SD = 0.44), the result was t(21) = 0.37, p = .356, and d = 0.08, indicating neither significance nor a substantial effect size.
A closer examination of the identification subscales unveiled a reduction in wishful identification within the similarity setting. For customization (PRE) (M = 1.8, SD = 0.82) and confrontation (POST) (M = 1.5, SD = 0.8), t(21) = 2.54, p = .01, and d = 0.54 indicated the overall strongest reduction across all subscales. Similarly, similarity identification also decreased significantly from customization (PRE) (M = 3.29, SD = 0.92) to confrontation (POST) (M = 3.05, SD = 1.09) with t(21) = 1.33, p = .034, and d = 0.41. In contrast, the overall smallest effect was observed for the embodied presence subscale in both settings. Upon direct comparison in terms of identification subscales, significant differences for each one were noted for customization (PRE) and confrontation (POST), except for wishful identification after the confrontation with the virtual agent (Figure \ref{pre}).

\begin{figure}[htbp]
\centerline{\includegraphics[width=10cm, height=6.2cm]{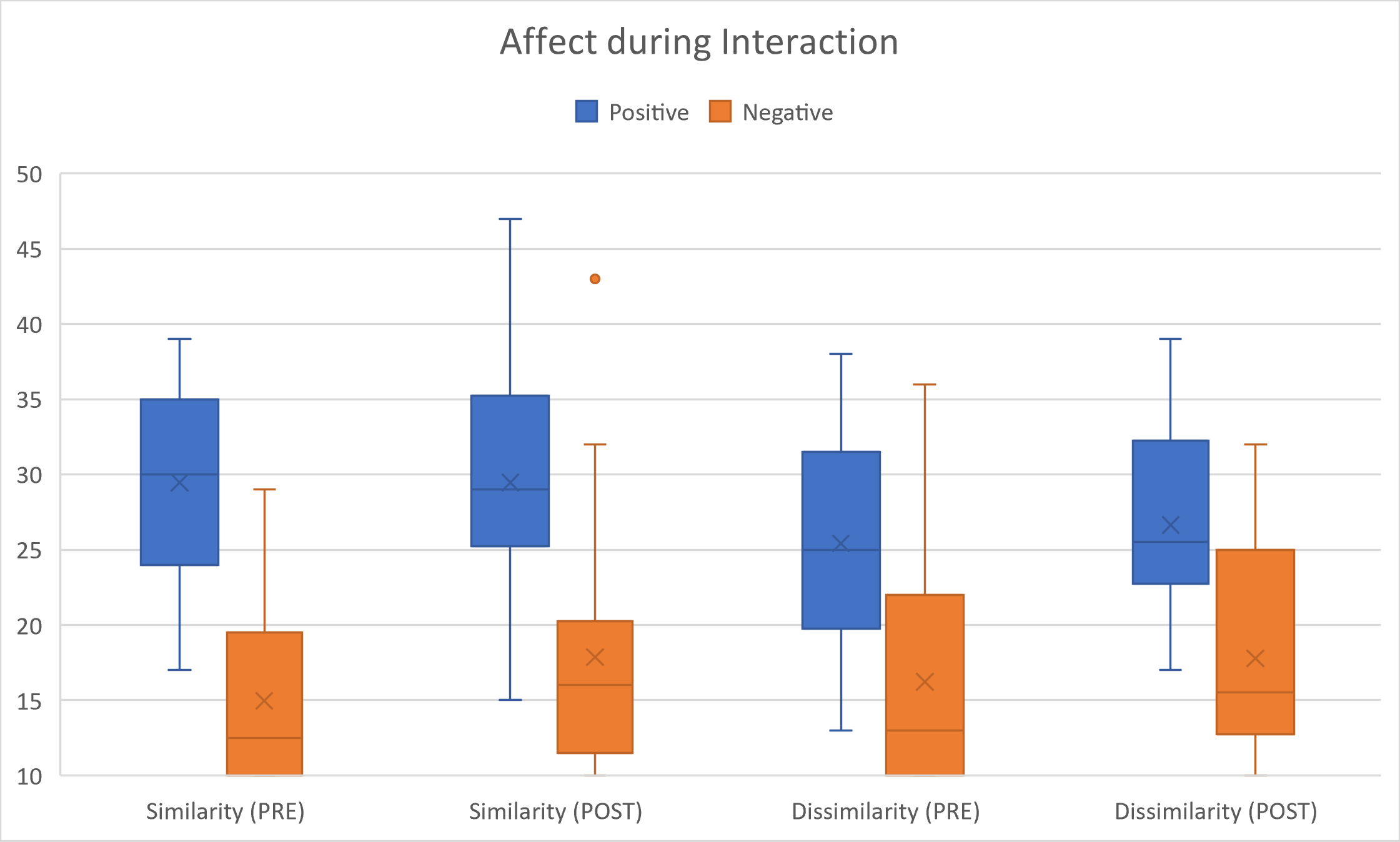}}
\centering
\caption{Positive and Negative Affect during Customization (PRE) and Confrontation (POST).}
\label{post}
\end{figure}

\textbf{H2:} As anticipated, the higher discrepancy between positive and negative affect in the similarity setting after the confrontation was confirmed to be overall higher compared to the dissimilarity setting. To validate this observation, the disparities between positive (M = 29.46, SD = 7.25) and negative affect (M = 17.86, SD = 8.1) in the similarity setting (POST) were contrasted with positive (M = 26.64, SD = 6.0) and negative affect (M = 17.77, SD = 6.61) in the dissimilarity setting (POST) was calculated. The t-test examining these differences resulted in t(21) = 3.02, p = .003, and d = 0.67, indicating a statistically significant distinction (Figure \ref{post}).

Upon closer examination, it was observed that a similar difference was found for customization (POST). The similarity setting (PRE) displayed a contrast in differences between positive (M = 29.45, SD = 6.0) and negative affect (M = 14.96, SD = 5.69) when compared to the dissimilarity setting (PRE) with positive (M = 25.41, SD = 6.48) and negative affect (M = 16.23, SD = 7.53). The t-test resulted in t(21) = 1.88, p = .037, and d = 0.4. It is noteworthy that positive affect in both settings is overall higher for the similarity setting, although only marginally significant. Another important finding is that the similarity setting exhibited a significant increase in negative affect from customization to confrontation. The comparison between customization (PRE) (M = 14.96, SD = 5.69) and after confrontation (POST) (M = 17.86, SD = 8.1) yielded t(21) = -1.82, p = .041, and d = 0.39.

\textbf{H3:} For this hypothesis, the negative impact of the confrontation on interest and enjoyment is notably more pronounced in the dissimilarity setting compared to the similarity setting, as corroborated in the following outcomes. In the similarity setting, the comparison of collected data for interest and enjoyment following customization (PRE) (M = 5.14, SD = 0.85) and confrontation (POST) (M = 4.73, SD = 1.1) yielded a t(21) = 1.91, p = .035, and d = 0.41. Conversely, in the dissimilarity setting, the comparison after customization (PRE) (M = 4.94, SD = 1.12) and confrontation (POST) (M = 4.42, SD = 1.03) resulted in t(21) = 2.28, p = .017, and d = 0.5. Therefore, statistical significance was observed in both settings in the before-after analysis, with the overall impact being higher in the case of the dissimilarity setting. Furthermore, no notable differences were observed when directly comparing the two settings before and after the confrontation.

\textbf{H4:} The subsequent hypothesis addresses the perceived value and usefulness of the confrontation, with the initial assumption that these factors would be significantly higher for the similarity setting (POST) (M = 4.41, SD = 1.58) compared to the dissimilarity setting (POST) (M = 4.22, SD = 1.5). However, the calculation following the confrontation yielded values of t(21) = 0.84, p = .206, and d = 0.18, indicating a lack of statistical significance. Consequently, the hypothesis is falsified. 
Upon closer examination of the collected data, the similarity setting exhibited a noteworthy increase when comparing customization (PRE) (M = 4.02, SD = 1.68) and confrontation (POST) (M = 4.41, SD = 1.58), resulting in t(21) = -1.95, p = .032, and d = 0.42, signifying statistical significance. In contrast, the dissimilarity setting did not yield similar results. Here as well, the direct comparison between the settings for both phases of the experiment shows no significant differences.

\begin{figure}[htbp]
\centerline{\includegraphics[width=10cm, height=6.2cm]{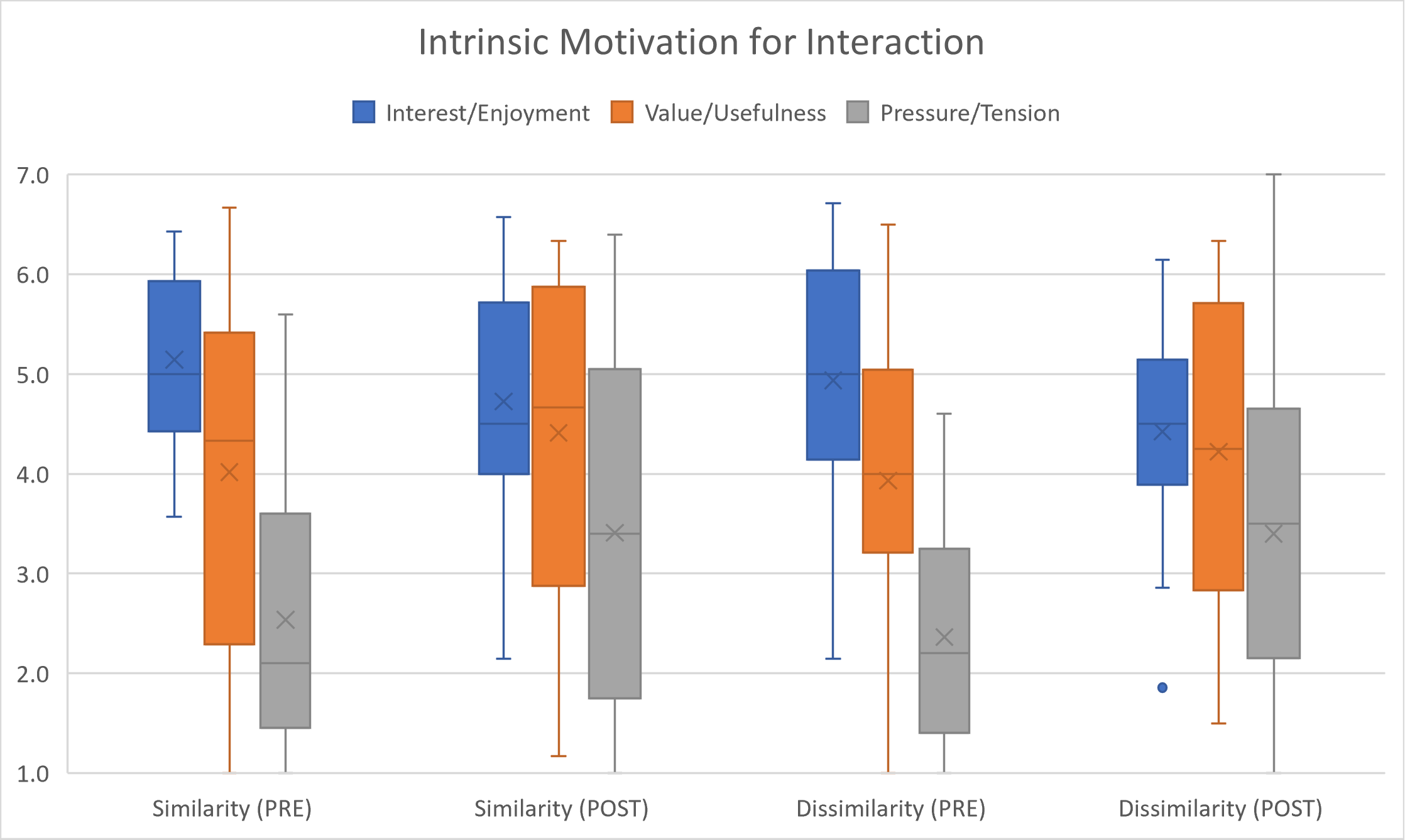}}
\centering
\caption{Intrinsic Motivation subscale ratings for interaction with Virtual Agent for both settings.}
\label{last}
\end{figure}

\textbf{H5:} Finally, the last hypothesis suggests that the perceived pressure and tension of each participant would increase to a greater extent in the dissimilarity setting, compared to the similarity one. This was verified using a t-test. In the similarity setting, after customization (PRE) (M = 2.54, SD = 1.35) and confrontation (POST) (M = 3.41, SD = 1.7), the values were t(21) = -2.92, p = .004, and d = 0.62. Conversely, for a dissimilar virtual agent, the comparison between customization (PRE) (M = 2.36, SD = 1.12) and confrontation (POST) (M = 3.4, SD = 1.52) resulted in t(21) = -3.6, p < .001, and d = 0.77. Thus, the change is significant in both cases, but the effect is more pronounced in the dissimilarity setting, thereby verifying the hypothesis (Figure \ref{last}).

\section{Discussion}

\subsection{Identification, Similarity, and Dissimilarity}
The results from the previous section suggest that confrontation, especially in the similarity setting, negatively affects identification. Notably, engagement with the virtual agent in a confrontation led to a general decrease in identification, potentially impacting positive effects like interest and enjoyment. However, this reduced identification could be desirable, aligning with the relevance of distancing oneself from negative beliefs.  

Also, in the similarity setting, similarity identification was significantly reduced. Surprisingly, this was not the case for the dissimilarity setting, in which it showed a slight increase. Although not significant, this raises questions about users suddenly perceiving the virtual agent as similar. Given minimal differences and the possibility of reporting errors, these cases leave open questions. Unexpectedly, high presence scores challenged the assumption that virtual agents' reduced interactivity would negatively impact presence. 

Overall. presence, along with similarity and wishful Identification, decreased post-confrontation, underlining the positive correlations among the subscales, as also mentioned in prior research\cite{hoffner2005, vanLooy2010}.

\subsection{Positive and Negative Affect}
Our second hypothesis suggested a greater difference between positive and negative affect would occur post-confrontation in the similarity setting compared to the dissimilarity setting. The results supported this hypothesis, showing that designing oneself might have been perceived as more satisfying than creating an arbitrary person. 
Although not conclusively proven in this study, it is reasonable to assume that self-design led to lower discrepancies and likely more positive feelings, as mentioned before. 
In the dissimilarity setting, users often deviated from their gender or chose amusing options. Thus, the dissimilar agent was likely distant from most users' self-concepts, possibly allowing more discomfort post-customization compared to the similarity setting. 

Significantly, negative affect increased post-confrontation in the similarity setting, while positive affect remained unchanged on average. While an increase in negative affect when confronted with unpleasant thoughts is understandable, a similar significant increase in positive affect could have also been expected if users successfully coped with the confrontation. Despite non-significant trends in positive affect increase in the dissimilarity setting, the overall higher mean values in the similarity setting suggest that confronting a self-similar agent was perceived as more unpleasant. Consideration of these findings, however, should be relative to dissimilarity setting values, which were inherently higher. 

\subsection{Intrinsic Motivation}
Although no significant differences in customization were observed, the dissimilarity setting's confrontation significantly decreased interest and enjoyment, aligning with the hypothesis these. A similar significant decrease was found in the similarity setting, where interest and enjoyment were higher on average. However, no significant differences emerged when comparing customization and confrontation values for similarity and dissimilarity.
The subsequent hypothesis addressed the perceived value and usefulness of confronting the virtual agent, suggesting that the similarity setting post-confrontation could potentially be perceived as more valuable. However, the data contradicted this hypothesis, indicating that both settings confrontation was perceived as useful, possibly because only the core exercise of contradicting the agent was considered. 
Lastly, tension and pressure during interaction were considered, with anticipation of a stronger increase in the dissimilarity setting. Although pressure increased significantly for both settings, it was stronger in the dissimilarity setting, aligning with the observed increase in negative affect. It is therefore reasonable to assume that confrontation does amplify negative feelings.

\subsection{Evaluation of the Therapeutic Tool}
Following the implementation of the therapeutic tool, supplementary data were gathered to rate the tool, the virtual agent, and the interaction with it. The tool garnered a positive reception overall, with a noticeable inclination towards the similarity setting. Participants expressed a preference for this setting, assigning significantly higher ratings compared to the dissimilarity setting. The general rating of the interaction with the virtual agent was positive, although evaluations of visual, auditory, and non-verbal behaviors received slightly lower ratings. The incorporation of a customization feature for the virtual agent within the framework of addressing dysfunctional cognitions was well-evaluated.

\subsection{Limitations}


In the course of the experiment, several limitations emerged that warrant consideration for the broader application of the results. Similarly, the study encounters challenges in establishing correlations between identification and outcomes. Although moderate correlations were discerned in the similarity setting, other significant dependencies remained either unattainable or only partially demonstrated. A proposition is made for a more stringent delimitation of variables and a larger sample size to potentially reveal explicit relations.

Moreover, the study's consideration of only external attributes for identification, owing to financial and temporal constraints, resulted in limited customization options. Participant feedback indicated a desire for a broader array of options encompassing hairstyles, facial textures, voices, and even features like age and height. While the available options were generally rated as sufficient, the absence of certain options might negatively impact user satisfaction.
Additionally, the study's one-week timeframe introduces a potential limitation in understanding shifts in identification levels over an extended period. Prolonged use may exert an influence on observed effects, and the procedural aspects of customization during extended use pose open questions. A suggestion was made to allow users the opportunity to modify the virtual agent over multiple sessions, considering the dynamic processes influencing self-perception. Temporary fluctuations in self-perception may occur due to factors such as mood or priming effects \cite{sedikides1992, demarree2005}. This approach aligns with the potential for shifts in identification levels over an extended timeframe, as noted in the literature \cite{klimmt2009, turkay2015}.

\section{Conclusion}

This study explored the impact of virtual agent similarity on user identification and the user experience in a therapy tool for treating dysfunctional cognition. Customization of self-similar and dissimilar agents was followed by a confrontation phase.
Results indicate higher identification with self-similar agents, especially in the similarity setting. Confrontation negatively affects identification, particularly in the similarity setting, with a significant reduction in wishful identification. 

Emotional experiences vary based on agent similarity, with self-similar settings correlating with higher positive affect and dissimilar settings displaying higher negative affect. 
Intrinsic motivation shows fluctuations post-confrontation, with increased interest during customization but a significant decrease in both settings.
Perceived value and usefulness increase post-confrontation, notably in the similarity setting. Elevated pressure and tension post-confrontation, also supported by increased negative affect, are more pronounced in the dissimilarity setting.

The study suggests potential links between identification and other factors. The visual appearance of a self-similar or dissimilar virtual agent significantly impacts emotional well-being and motivation, potentially influencing therapy success.
Identification, especially when supplemented by customization, may enhance motivation for prolonged interaction. The study provides insights into customization, virtual agent identification, and their implications in therapy. Validation and further investigation into negative emotional experiences are recommended.

%
%
 \bibliographystyle{splncs04}
 \bibliography{samplepaper.bib}
%

\end{document}